\documentclass[journal]{IEEEtran}
\ifCLASSINFOpdf
\else
\fi

\usepackage{lineno,hyperref}
\usepackage{amsmath,amssymb,amsfonts}

\usepackage{xcolor}

\usepackage{subfigure}
\usepackage{graphicx} 

\usepackage{mathtools} 

\usepackage{textcomp}

\usepackage{soul}

\usepackage{algorithm}

\usepackage[noend]{algorithmic}

\DeclareMathOperator*{\Max}{Max}

\begin{document}
\title{A Generalized Framework for Joint Dynamic Optimal RF Interface Setting and Next-Hop Selection in IoT networks with Similar Requests}

%
%
%

\author{Monireh~Allah~Gholi~Ghasri,
			Ali~Mohammad~Afshin~Hemmatyar
\thanks{A. M. A. Hemmatyar was with the Department of Computer Engineering, Sharif University of Technology, Azadi Ave., Tehran, Iran,
 e-mail: (hemmatyar@sharif.edu).}
\thanks{Manuscript received .... .., 2021; revised .... .., 2021.}}

%
%

\markboth{Journal of IEEE Networking Letters,~Vol.~.., No.~.., .....~2021}
{A. G. Ghasri and Hemmatyar: A Generalized Framework for Joint Dynamic Optimal RF Interface Setting and Next-Hop Selection in IoT networks with Similar Requests}
\maketitle

\begin{abstract}
Many applications which run on the Internet-of-Things machines require similar bandwidths. These machines are equipped with multiple Radio Frequency(RF) interfaces for machine-to-machine or machine-to-Base Station(BS) communications. Joint dynamic optimal RF interface setting and next-hop selection in data transmission can maximize the network data rate. For this purpose, we propose a generalized framework for simultaneous selection of these components according to the required bandwidth of machines. This framework is suitable for networks with multiple BSs and sources with similar bandwidth requests. Simulation results show that using the proposed framework, the average data rate of network sources increases up to 117\%.
\end{abstract}

%
\IEEEpeerreviewmaketitle

\section{Introduction}
\IEEEPARstart{N}{owadays}, Internet-of-Things (IoT) devices with various applications need different requirements, such as BandWidth (BW) \cite{bMSAWiLt2012}. On the other hand, these devices (machines and Base Stations (BS)) are equipped with multiple Radio Frequency (RF) interfaces, such as Z-Wave, Bluetooth, and WiFi for Machine-to-Machine communication (M2M) and NB-IoT, LTE-M, and LTE for Machine-to-BS (M2B) communication. Therefore, they can use these RFs according to their needs. Optimal RF selection according to the required BW of the network sources maximizes their total data rate. Simultaneous use and dynamic optimal selection of these RF interfaces along with next-hop (relay or BS) selection helps to improve network data rate and reduce the interference of communication signals in practice \cite{HMSFCaSi2021, MRSRGhHe2020, AMHCHuXu2017, PORISiJK2016}. Using the word \textbf{dynamic} in the literature on RF interface setting means that instead of using only constant RFs statically in communications, similar to an optimal relay selection by static RF interfaces previously provided \cite{MRSRGhHe2020}, RF should be dynamically selected from the existing RFs according to the need.

Despite more attention to the proper selection of relays to improve network coverage and data rate \cite{RSPAKhMa2020} where the direct channel with the BS is weak or disconnected in previous works, the importance of RF interfaces setting especially when choosing the next-hop in transmission has received less attention. In this regard, while setting up the RF interface used in smartphones is done manually or automatically by default, some studies provide better methods by appropriate link and time determination or simultaneous usage of multiple RF interfaces for handover\cite{ HMSFCaSi2021, WLHMBerg2019}. Some other works in this area include maximizing the multi-cell unlicensed LTE and WiFi networks throughput \cite{JRSNWuHe2021}, and improving throughput and delay by concurrent transmission with dual connectivity \cite{DODCHeHu2021}. Unlike previous studies that focused more on dual-connectivity in M2B communication, in a recent work, a dynamic optimal RF interface setting for M2M communication with relay selection, called DORSA, in a network cell with a BS has been considered \cite{DORSGhHe2021}.

To complete these approaches, a comprehensive framework is provided for optimally selecting from M2M‌ and M2B RF interfaces, along with the next-hop of data transmission. Unlike mentioned works \cite{HMSFCaSi2021, MRSRGhHe2020, RSPAKhMa2020, WLHMBerg2019, JRSNWuHe2021, DODCHeHu2021, DORSGhHe2021}, in the proposed framework, the dynamic optimal selection between both types of M2M RFs and M2B RFs(for capillary and cellular part of a hybrid network, respectively) is done simultaneously with each other and with relay selection. In addition, unlike some previous work \cite{MRSRGhHe2020, DORSGhHe2021}, there is no limit to the Number of (No.) BSs in the proposed problem-solving framework. Therefore, this paper can be an extension of some previous works \cite{MRSRGhHe2020, DORSGhHe2021}.

\textbf{Main Contributions:} In this paper, the following contributions are provided:
\\- A new generalized algorithmic framework for optimal selection of RF interfaces and the next-hop (relay or BS) for data transmission according to the requested BW of sources in an IoT network is proposed.
\\-Our framework can support IoT networks including multiple BSs and machines, each of which is equipped with multi-RF interfaces.
\\- In our framework, all of the machines can use different multiple M2M and M2B RF equipment (for capillary and cellular part of the network, respectively) simultaneously to send data of sources with the same requested BW.
\\- Finally, using the presented framework can help to detect the upper bound of the average total data rate of network sources to examine other algorithms for next-hop selection and RF interface setting.

\section{System Model}

\begin{figure}[!htb]
\centering
\includegraphics[scale=0.4]{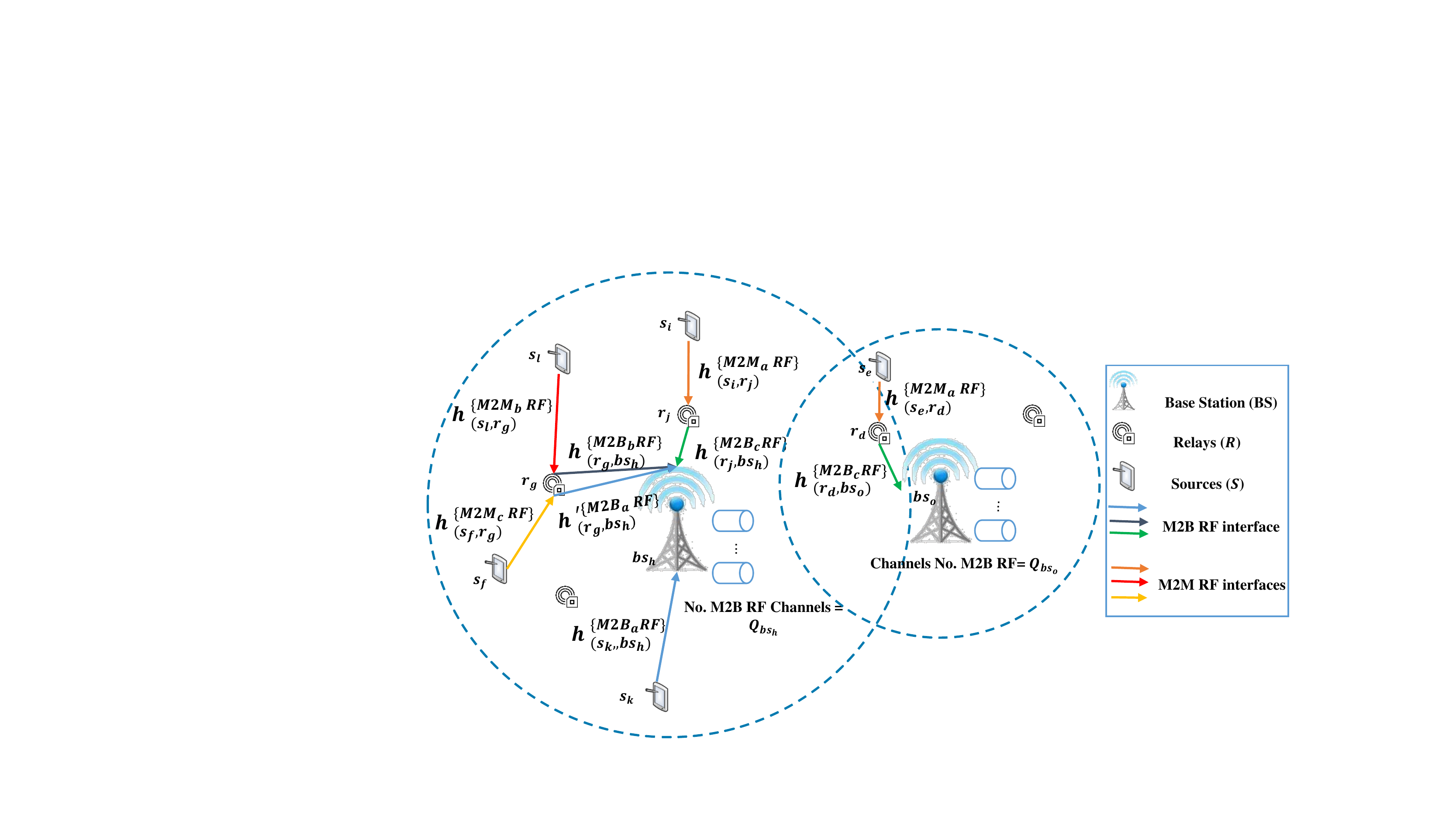}
\caption[Illustration of devices and communications in the network model.]{Illustration of devices and communications in the network model.}
\label{fig:myScenario}
\end{figure}

In this model, an IoT network with $N$ fixed machines and $N_b$ BSs is considered. Each machine is equipped with $t^{M2M}$ M2M RF interfaces and $t^{M2B}$ M2B RF interfaces and each BS is equipped with $t^{M2B}$ M2B RF interfaces. Among the machines, there are $N_s$ sources and $N_r$ relays that these $N$ machines, where $N=N_s+N_r$, are randomly placed around BSs with a uniform distribution. An illustration of devices and communications in the network model is shown in Fig.~\ref{fig:myScenario}.

\textbf{Assumptions:} The assumptions for providing the desired framework are as follows:
\begin{enumerate}
\item The relation between sources with relays and base stations is N-to-One.
\item The No. different channels of each M2M RF interface is equal to one.
\item Different M2M and M2B RF interfaces do not overlap and do not interfere with each other (such as WiFi with Bluetooth, LTE, and etc.). 
\item Requested BW of all sources are the same.
\item The relaying protocol is Decode-and-Forward (DF).
\end{enumerate}
\textbf{Problem Formulation:} Here are some of the relations used to formulate the problem. The calculated data rate of the communication by DF relaying between source $s$ and BS $b$ through relay $r$ is estimated by relation (\ref{eq:Capsrd}) in a time slot \cite{IRSTZhXu2013}.
%
\begin{equation}
\label{eq:Capsrd}
	C_{{s,b}_{through r}} = min \lbrace C^{ t^{M2M} }_{s,r}, C^{  t^{M2B}  }_{r,b} \rbrace,
\end{equation}.
where $C^{ t^{M2M}  }_{s,r}$ is maximum data rate between $s$ and $r$ on $t^{M2M}$th RF interface and $ C^{  t^{M2B}  }_{r,b}$ is maximum data rate between $r$ and $b$ on $t^{M2B}$th RF interface. 
To calculate the maximum data rate between two devices (machine or BS) using $t$th RF ($C^{ t^{M2M|B}}_{(n_i, n_j)}$), we consider the minimum value between the data rate calculated for the channel (regardless of $t$th RF capacity) and the maximum data rate capacity for $t$th RF ($C^{ t^{M2M|B} }_{Max} $).
\begin{equation}
\label{eq:Capsrd_t}
	C^{  t^{M2M|B}  }_{(n_i, n_j)} = min \lbrace C^{  t^{M2M|B}  }_{(n_i, n_j)-SH} , C^{ t^{M2M|B} }_{Max} \rbrace,
\end{equation}.
where $C^{  t^{M2M|B}  }_{(n_i, n_j)-SH}$ is the maximum direct channel data rate between $i$th and $j$th nodes on $t^{M2M|B}$th RF interface. It is calculated using the Shannon-Hartley (SH) theorem from relation (\ref{eq:Capninj}):
\begin{equation}
\label{eq:Capninj}
	C^{  t^{M2M|B}  }_{(n_i, n_j)-SH} = B^{   t^{M2M|B}  }_{(n_i, n_j)}   log_2(1+\mathrm{SINR^{   t^{M2M|B}  }_{(n_i, n_j)}}),
\end{equation}	
where $B^{   t^{M2M|B}  }_{(n_i, n_j)}$ is the BW of $t^{M2M|B}$th RF interface channel between $i$th and $j$th nodes, and $SINR^{   t^{M2M|B}  }_{(n_i, n_j)}$ is Signal-to-Interference-plus-Noise-Ratio (SINR) of $t^{M2M|B}$th RF interface channel between $i$th and $j$th nodes.
Now, relation (\ref{eq:optEq_srRFb}) formulates the optimal next-hop selection and dynamic RF interface setting with the aim of maximizing the total data rate of network sources.
\begin{align}
\label{eq:optEq_srRFb}
    &\Max_{ x, y, z}
    \begin{aligned}[t]
       &\sum_{i=0}^{(N_s-1)}{\sum_{j=0}^{(N_q-1)}\sum_{k=0}^{N{p} -1}  } \\
       &  x_{i,j} y_{j,k} c^{'''}_{i,k} + \sum_{i=0}^{(N_s-1)}\sum_{k=0}^{(N_p -1)} z_{i,k} c^{'}_{i,k},  \\
    \end{aligned} \notag \\\\  
        \text{Subject to} \notag \\
    & x_{i,j},y_{j,k},z_{i,k} \in \lbrace 0, 1 \rbrace: \notag\\   
    &\forall i,j,k:0 \leq i < N_s , 0 \leq j < N_q, 0 \leq k <  N_p, \label{eq:optEq_xijyjkzik}\\   
    & \sum_{i=0}^{(N_s-1)} x_{i,j} \leq 1 , \sum_{k=0}^{(N_p -1)} y_{j,k} \leq 1, \forall j:0 \leq j < N_q, \notag \\
 \label{eq:optEq_firstnEq}\\
    & \sum_{j=0}^{(N_q -1)} x_{i,j} \leq 1 , \notag \\
     & \sum_{k=0}^{(N_p -1)} z_{i,k} \leq 1, \forall i:0 \leq i < N_s), \label{eq:optEq_secnEq}\\    
& \sum_{i=0}^{(N_s-1)}{\sum_{j=0}^{(N_q-1)}\sum_{k=0}^{(N_p -1)}  x_{i,j} y_{j,k}} \notag \\ 
& + \sum_{i=0}^{(N_s-1)}\sum_{k=0}^{(N_p -1)} z_{i,k} \leq N_b Q_{BS} \label{eq:optEq_thirthnEq}.
\end{align}
The used variables and problem constraints are defined as follows:
\\-The No. $Rel-RF^{M2M}-BS-RF^{M2B}$ quadruple and $BS-RF^{M2B}$ pair are represented by $N_q = (N^{M2M}_t  N^{M2B}_t N_r N_b)$ and $N_p=N^{M2B}_t N_b$, respectively.
\\- $x_{i,j}$: is 1 if $i$th source has selected the $j$th $Rel-RF^{M2M}-BS-RF^{M2B}$ quadruple and 0 otherwise,
\\- $y_{j,k}$: is 1 if $j$th $Rel-RF^{M2M}-BS-RF^{M2B}$ quadruple has selected the $k$th $BS-RF^{M2B}$ pair and 0 otherwise,
\\- $z_{i,k}$ : is 1 if $i$th source has selected the $k$th $BS-RF^{M2B}$ pair and 0 otherwise,
\\- $c_{i,j}$: the data rate between $i$th source and $j$th $Rel-RF^{M2M}-BS-RF^{M2B}$ quadruple,
\\- $ c^{'}_{i,k}$: the data rate between $i$th source and $k$th $BS-RF^{M2B}$ pair,
\\- $c^{''}_{j,k}$: the data rate between $j$th $Rel-RF^{M2M}-BS-RF^{M2B}$ quadruple and $k$th $BS-RF^{M2B}$ pair,
\\- $c^{'''}_{i,k} = min \lceil c_{i,j}, c^{''}_{j,k} \rbrace$ (according to relation (\ref{eq:Capsrd})),
\\- $N_s$: the No. sources. 
\\- $N_r$: the No. relays,
\\- $N^{M2M}_t$: the No. M2M RF interfaces,
\\- $N^{M2B}_t$: the No. M2B RF interfaces,
\\- $Q_{BS}$: quota or total connection capacity of each BS which is obtained by $Q_{BS} = \sum_{t^{M2B}=0}^{(N_{t}^{M2B} -1)} Q^{t^{M2B}}_{BS}$, where $Q^{t^{M2B}}_{BS}=  min \lbrace  {\lfloor  \frac{BW^{t^{M2B}}_{BS}}{BW_s} \rfloor},{Cap^{t^{M2B}}_{BS}} \rbrace$, $BW^{t^{M2B}}_{BS}$ is the BW of the M2B RF interface, $BW_s$ equals to requested BW of sources, and $Cap^{t^{M2B}}_{BS}$ is the capacity of the No. user connections to $t^{M2B}$'th RF.
\\- The first summation of inequality (\ref{eq:optEq_firstnEq}) represents each $Rel-RF^{M2M}-BS-RF^{M2B}$ quadruple can only be assigned to a single source.
\\- The constraint of the second summation in inequality (\ref{eq:optEq_firstnEq}) is that each $Rel-RF^{M2M}-BS-RF^{M2B}$ quadruple can only be connected to a single BS is represented.
\\- The constraint of the first summation in inequality (\ref{eq:optEq_secnEq}) is that each source can only be connected to a single $Rel-RF^{M2M}-BS-RF^{M2B}$ quadruple.
\\- The constraint of the second summation in inequality (\ref{eq:optEq_secnEq}) is that each source can only be connected to a single $BS-RF^{M2B}$ pair.
\\- The constraint of the summation in inequality (\ref{eq:optEq_thirthnEq}) represents the total No. two-hop (the first summation) and direct (the second summation) connections to BSs is less than or equal to the total connection capacity of BSs ($N_b Q_BS$).

\section{The Proposed Framework for Dynamic Optimal Next-Hop Selection and RF Interface Setting Algorithm (DONSA)}
\begin{figure}[!htb]
\centering
\includegraphics[scale=0.29]{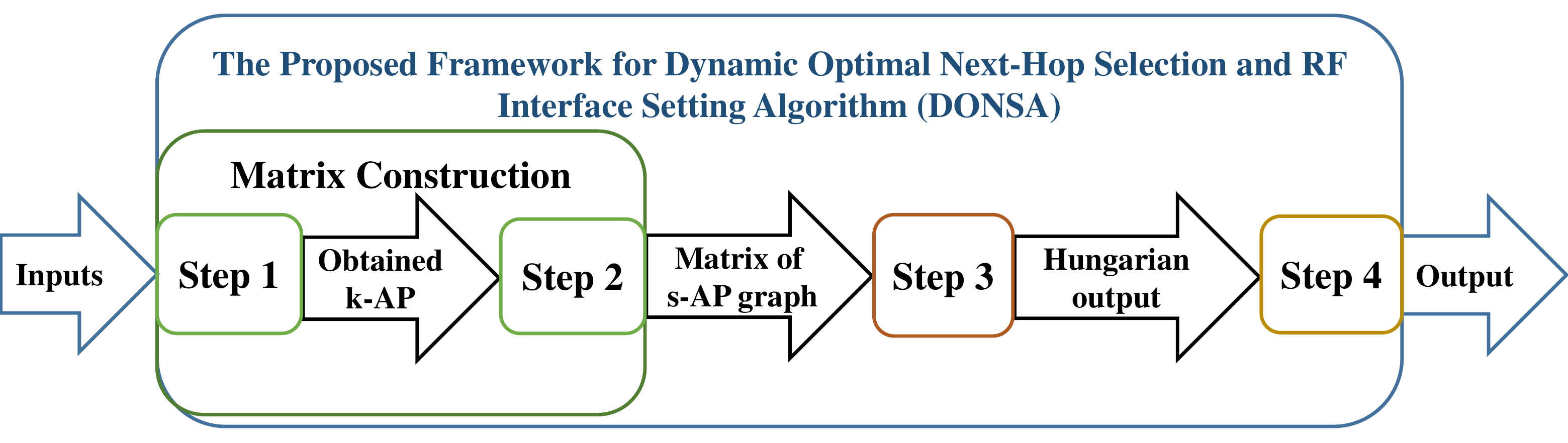}
\caption[Schematic of the proposed framework for DONSA.]{Schematic of the proposed framework for DONSA.}
\label{fig:ourFrameworkScheme}
\end{figure}

%
To generalize DORSA \cite{DORSGhHe2021}, we provide a generalized framework for the RF interface and next-hop selection algorithm in an IoT network, with no restrictions on RF interface type (M2M or M2B), No. RF interfaces, No. machines, and No. BSs. The following set of steps will be implemented under the mentioned generalized framework. The No. network components (machines, BSs, and RF interfaces) and the data rate between devices are inputs of the desired framework. A schematic of the proposed framework for DONSA is shown in Fig.~\ref{fig:ourFrameworkScheme}. In the following, we explain the steps of solving the problem using a $k$-AP solver \cite{MRSRGhHe2020} in this framework with a brief explanation. The optimality of the $k$-AP solver used in our proposed algorithm has already been proven \cite{MRSRGhHe2020}.

\textbf{Step 1-} \textit{Transforming dynamic optimal RF interface setting and next-hop selection problem to a $k$-AP:} To solve the desired problem, firstly we must model the problem into a $k$-AP. To achieve this model, a weighted bipartite graph equivalent to the principal problem is defined. In this graph, we seek to maximize the total weight of the selected edges. 

Now, we define the vertices of the graph. One part of the graph (e.g. the left side) contains vertices equivalent to the sources (as in previous works \cite{MRSRGhHe2020, DORSGhHe2021}), but the other part dedicated to this problem must be defined in such a way that the definition of the desired problem does not change and there should be one-to-one connections between the two parts.

For this purpose, after studies, we define $Rel-RF^{M2M}-BS-RF^{M2B}$ quadruple and $BS-RF^{M2B}$ pair as vertices representing the elements of the second part (e.g. the right side).
The maximum No. possible edges of the bipartite graph is equal to $k = min (N_s, N_bQ_b)$, where the connection capacity of each BS is equal to $Q_b = \sum^{N_{t^{M2B}}-1}_{l=0} N_{ch_{{t^{M2B}}_l}}$, and $N_{ch_{{t^{M2B}}_l}}$ is the No. channels of each M2B RF interface. Thus, the main problem transforms to a $K$-AP.

\begin{figure}[!htb]
\centering
\includegraphics[scale=0.45]{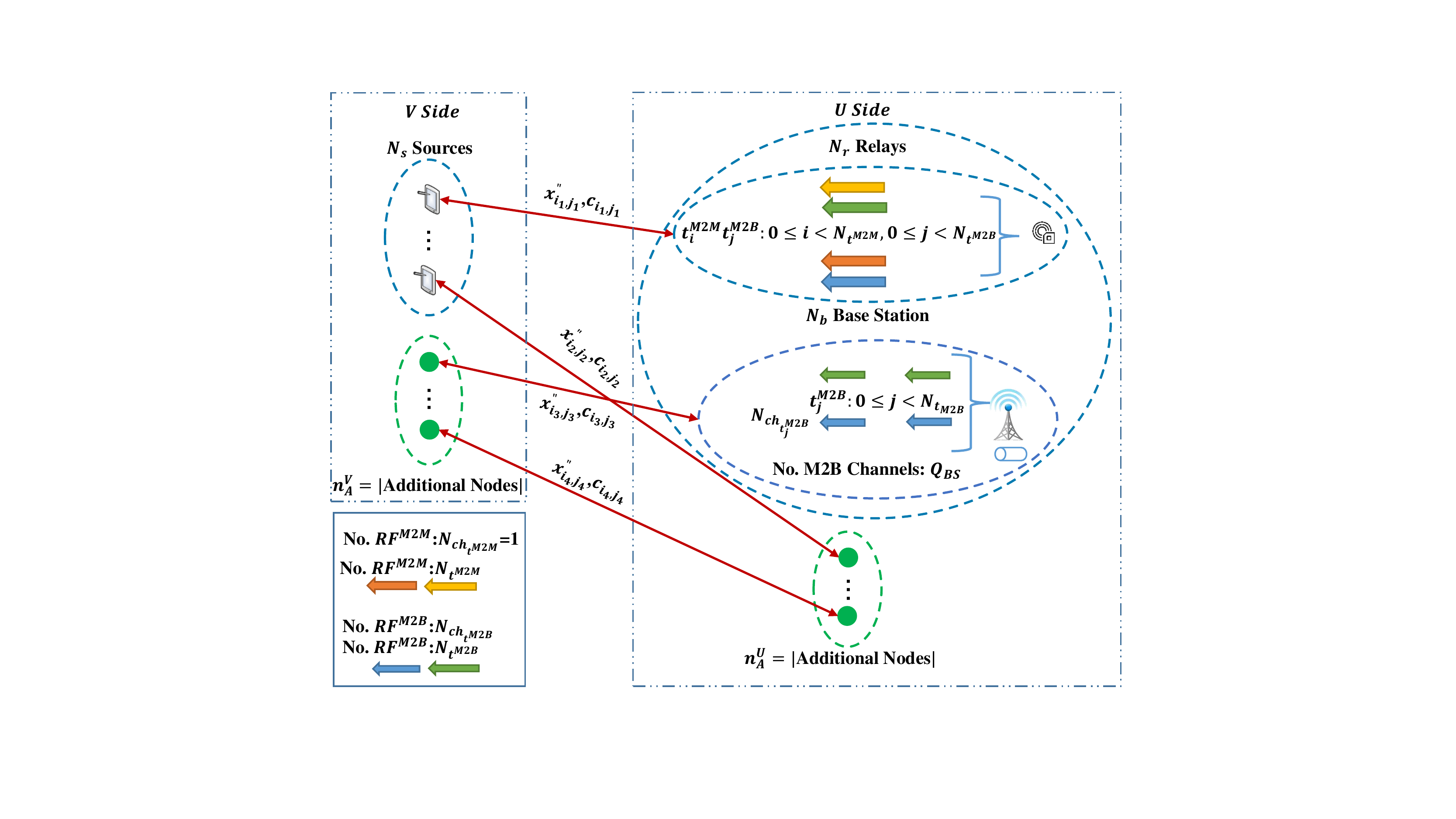}
\caption[Schematic of the obtained s-AP graph in step 2 of DONSA.]{Schematic of the obtained s-AP graph in step 2 of DONSA.}
\label{fig:ourFrameworkGraph}
\end{figure}

\textbf{Step 2-} \textit{Transforming the obtained $k$-AP to an standard assignment problem (s-AP):} Now to remove the $k$ edge selection constraint in the graph of $k$-AP, the obtained $k$-AP is transformed to an s-AP. For this purpose, we must transform the graph in such a way that there is no constraint to the selection of the No. edges \cite{ MRSRGhHe2020, DORSGhHe2021}. Therefore, without losing the generality of the problem, we add a No. vertices and related edges to one or both parts of the graph. Now, in this way, we can reach the equivalent answer to the main problem in the new s-AP, with the aim of maximizing the total weight of the selected edges in the new graph. The schematic of the obtained s-AP graph in step 2 of DONSA is shown in Fig.~\ref{fig:ourFrameworkGraph}.

Without losing generality, the weight of edges with at least one new additional vertex is defined as follows:
\\- The weight of edge with the second vertex from new vertices on the other part is equal to zero which has no effect on the selection of edges, and 
\\- The weight of edge with the second vertex from the previous vertices on the other part is equal to infinite (or in practice large enough $A_{value}$, for example $A_{value}=1 + \sum_{edges} wight^{Any Edge}$).

Also, in general, the No. additional vertices are as follows:
\begin{enumerate}
\item if $N_s \geq N_b Q_{BS}$: $N_s-k$ vertices are added to the right side and $N^{M2M}_t N^{M2B}_t N_r N_b$ vertices are added to the left side to cover unmatched sources.

\item else if $N_s < N_b Q_{BS}$: Due to the fact that the connection capacity of BSs is more than the No. sources and all sources can be matched, there is no need to add a new vertex to the right side of the graph. Just to make the two sides of the graph symmetrical, equivalent to the No. differences between the No. vertices on the right side and the left side (e.g. $N^{M2M}_t N^{M2B}_t N_r+N_b Q_{BS}-N_s$), add the vertex to the left side.


\end{enumerate}



Note: if No. channels per each M2B RF interface ($N^{M2B}_{{ch}_t}$) was more than $N_s$, initially, the maximum $N^{M2B}_{{ch}_t}$ can be considered equal to $N_s$ to reduce the time complexity of the problem solution.

\textbf{Step 3-} \textit{Solving the obtained s-AP:} In some studies, to solve the optimal relay selection problem, it can be transformed to an s-AP, then the available tools such as the Hungarian algorithm can be used to solve it \cite{MRSRGhHe2020, DORSGhHe2021}. Since the Hungarian algorithm is a common solver with a polynomial-time complexity, that is able to find the optimal solution of an s-AP as a maximum weighted matching problem \cite{ICRDchitr2016}, in the proposed framework, it is used to solve the obtained s-AP. 

\begin{figure}[!htb]
\centering
\includegraphics[scale=0.3]{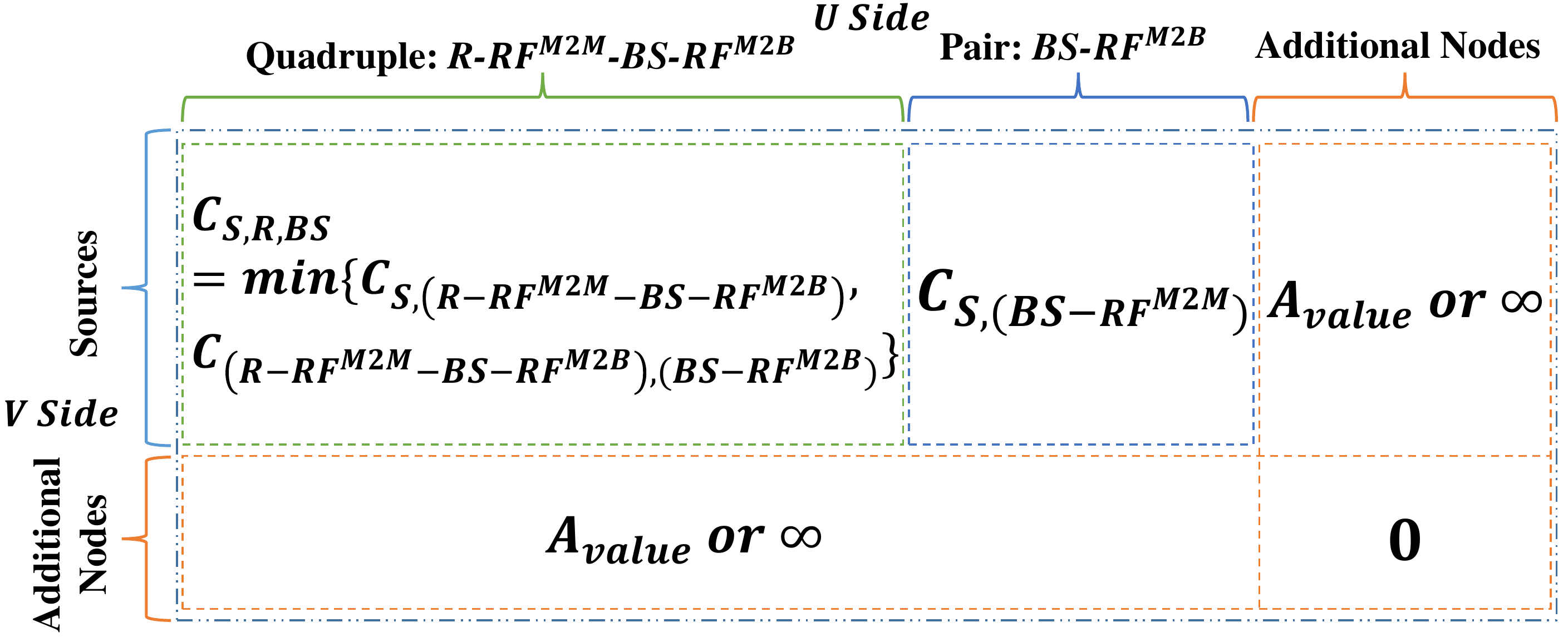}
\caption[The equivalent matrix of the s-AP graph in step 2 of DONSA.]{The equivalent matrix of the s-AP graph in step 2 of DONSA.}
\label{fig:ourFrameworkMatrix}
\end{figure}

 Now, in this step, the equivalent matrix to the obtained s-AP graph in step 2 of DONSA is given as the input of the Hungarian algorithm. In the other words, the elements of this matrix represent the connection between network devices and the data rate of one-hop and two-hop connections between sources and other network devices (relays and BSs) on different RFs. Fig.~\ref{fig:ourFrameworkMatrix} shows a view of the equivalent matrix of the s-AP graph in step 2 of DONSA, which is one of the main elements of this framework. 
 
\textbf{Step 4-} \textit{Obtaining the final results of the desired problem:} Let us now consider the first $N_s$ elements of the output of the Hungarian algorithm, which are related to the final response for the sources. If the output content associated with a source is from new vertices, it means that in the optimal solution of the desired problem, this source will not be connected to any node of the network. For other sources, in the final optimal assignment, each source whose output element content corresponds to:
\\- A $(Rel-RF^{M2M}-BS-RF^{M2B})_{output}$ quadruple means that the source must use the specified RF interfaces to send data to the intended BS via the specified Rel,
\\- A $(BS-RF^{M2B})_{output}$ pair means that the source must use the specified RF interface to send data directly to the intended BS. 
Algorithm \ref{alg:DONSAalg} presents the pseudo code of DONSA.
\begin{algorithm}[!htb] 
\caption{Pseudo code of DONSA}
\label{alg:DONSAalg}
\begin{algorithmic}[1]
\algsetup{linenosize=\small}
\scriptsize
		\STATE \textbf{Inputs:} No. machines, No. BSs, No. M2M and M2B RF interfaces, and the data rate between devices,
        \STATE \textbf{Step 1:} 1’th Transformation (the selection problem to a k-AP)$\Rightarrow$Output: Obtained k-AP, 
        \STATE \textbf{Step 2:} 2’th Transformation (k-AP to a s-AP)$\Rightarrow$Output: Equivalent matrix to the s-AP graph,
        \STATE \textbf{Step 3:} Solving the obtained s-AP by Hungarian algorithm$\Rightarrow$Output: The output vector of the Hungarian algorithm, 
        \STATE \textbf{Step 4:} Obtaining the final results$\Rightarrow$Output: Final assignment. 
\end{algorithmic}
\end{algorithm}
\textbf{Time Complexity (TC)}: Examining TC of steps, as previous works \cite{MRSRGhHe2020, DORSGhHe2021}, it is observed that the bottleneck is the Hungarian algorithm that if its input matrix size$=n \times n$ then TC$= O(n^3)$ \cite{ICRDchitr2016}. Therefore, TC of DONSA is equal to $O(n^3)$, where $n = N_s+N^{M2M}_t N^{M2B}_t N_r N_b$, if $N_s \geq N_b Q_{BS}$,and otherwise $n = N^{M2M}_t  N^{M2B}_t  N_r N_b+ N_bQ_{BS}$.
\section{Simulation Results}
The simulations were implemented in M2MSim \cite{DORSGhHe2021} with 200 runs on a device with a 4-core Intel Xeon and 4 GB of RAM. An uplink network cell with default size $500 \times 500$ and 1 BS in the middle of the cell, $N^{M2M}_t=3$ M2M RF interfaces, Z-Wave, Bluetooth, and WiFi  (as a broadband RF with more BW than other M2M RF interfaces under consideration), and $N^{M2B}_t=3$ M2B RF interfaces, NBIoT, LTE-M, and LTE (as a broadband RF with more BW than other M2B RF interfaces under consideration), is simulated. Also, path loss exponent $\beta=4$, shadowing (dB) and small scale fading (dB) are modeled by $\mathcal{N}(0,64)$ and $Rayleigh(1)$, respectively. In this paper, Average Data Rate (ADR) of sources, No. Unmatched Sources (NUS), and Average Execution Time (AET) of the generalized framework of DONSA using WiFi, Bluetooth, and Z-Wave as M2M RFs and LTE, LTE-M, and NBIoT as M2B RFs (\textbf{DONSA\_WBZ-LMN}) is compared to the following algorithms in 3 scenarios: \textbf{DiTOSA\_L}\cite{DORSGhHe2021}: the optimal direct transmission to BS with LTE, \textbf{SORSA\_W-L}\cite{MRSRGhHe2020}: the optimal relay selection with static M2M (WiFi) and M2B (LTE) RF interface setting, and \textbf{DORSA\_WBZ-L}\cite{DORSGhHe2021}: DORSA with WiFi, Bluetooth, and Z-Wave as M2M RFs and LTE as M2B RF.

\begin{figure}[!htb]
    \centering
    \subfigure[Scenario 1: No.machines=300, req.BW=200(kHz), cell radius=500(m).]
    {
        \includegraphics[scale=0.45]{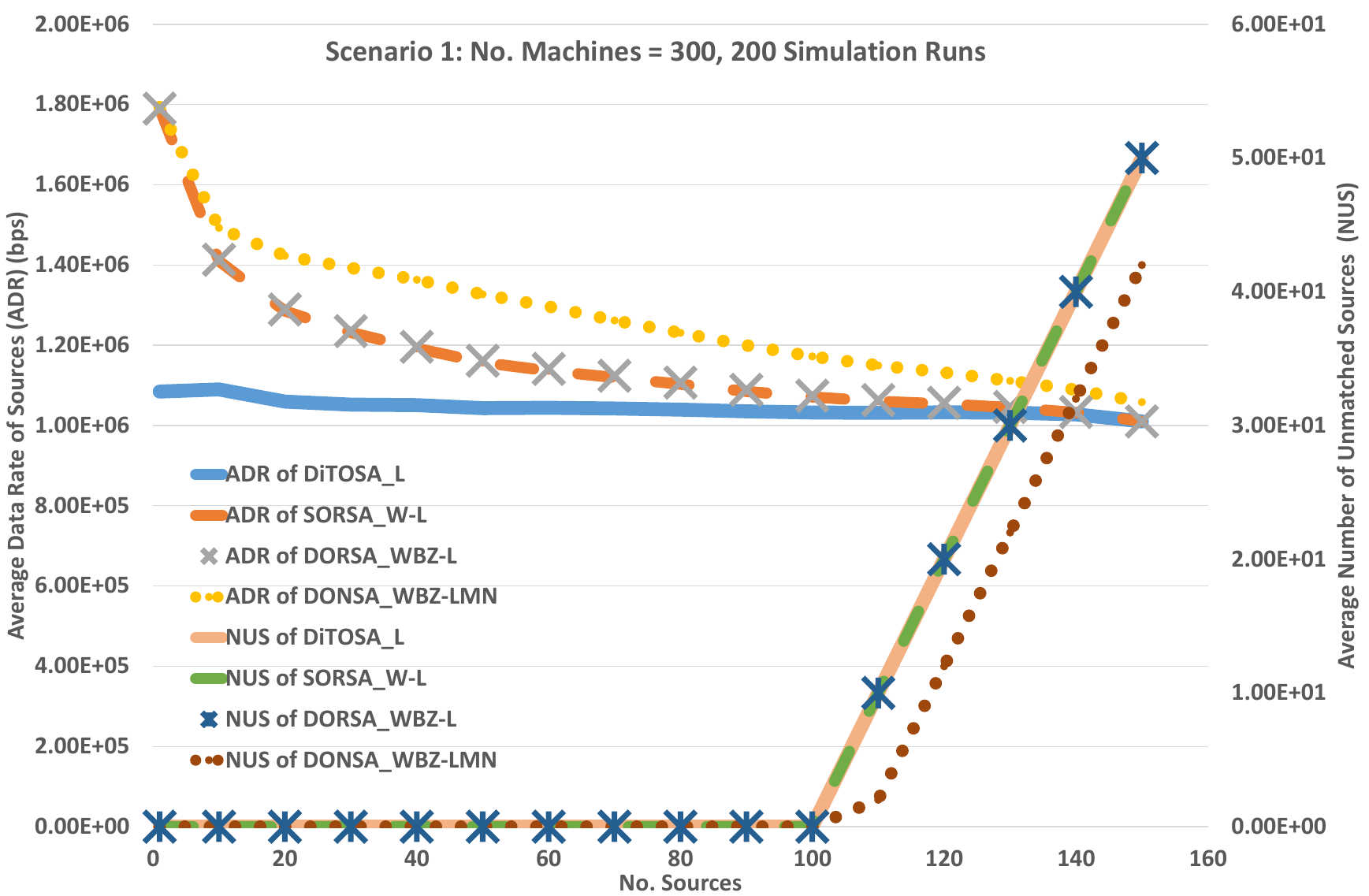}
        \label{fig:s1_ADR_NUS}
    }
    \subfigure[Scenario 2: No.sources=150, No.relays=150, req.BW=200(kHz).]
    {
        \includegraphics[scale=0.45]{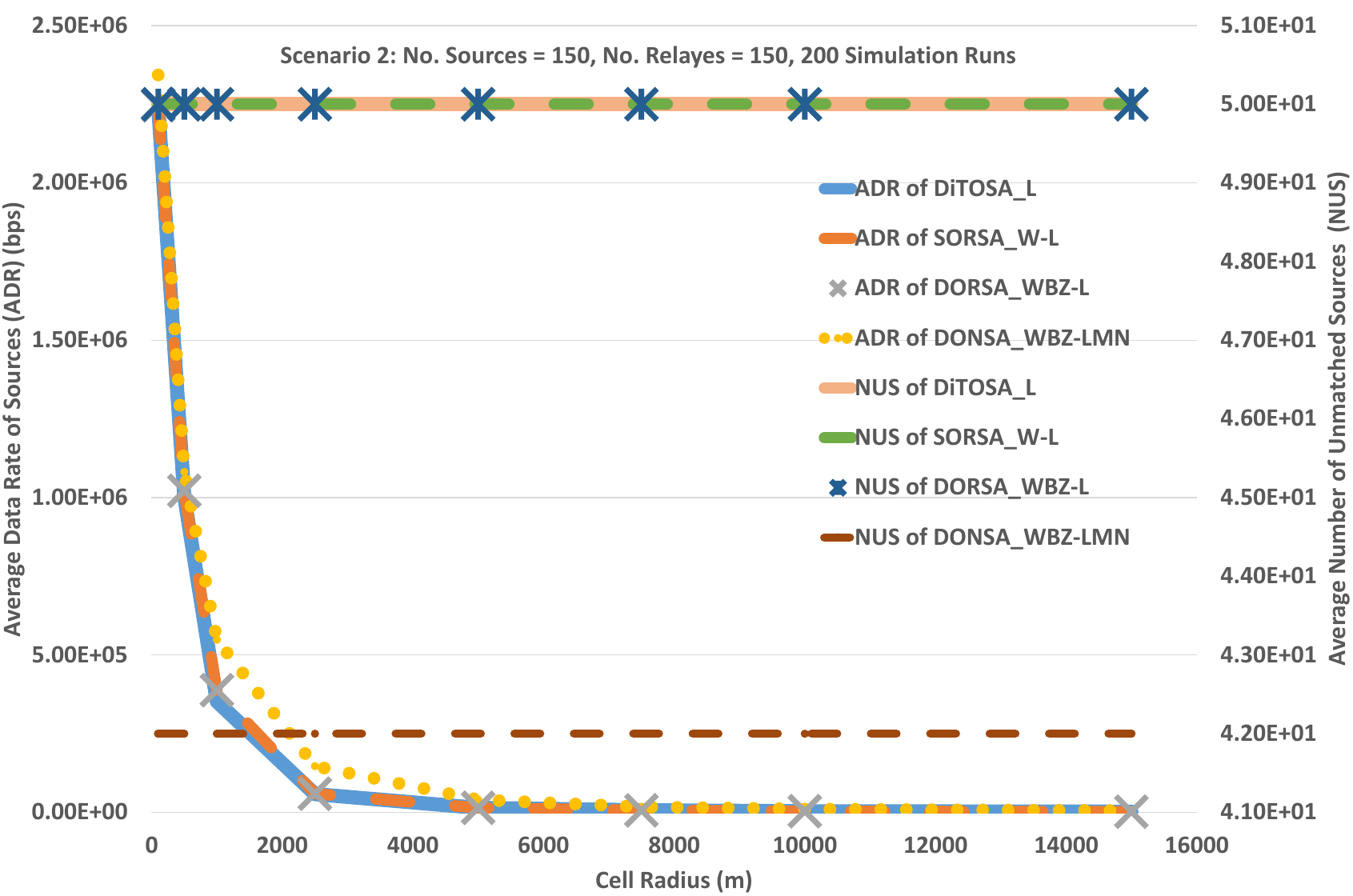}
        \label{fig:s4-2_ADR_NUS}
    }
    \subfigure[Scenario 3: No.sources=150, No.relays=150, cell radius=500(m).]
    {
        \includegraphics[scale=0.45]{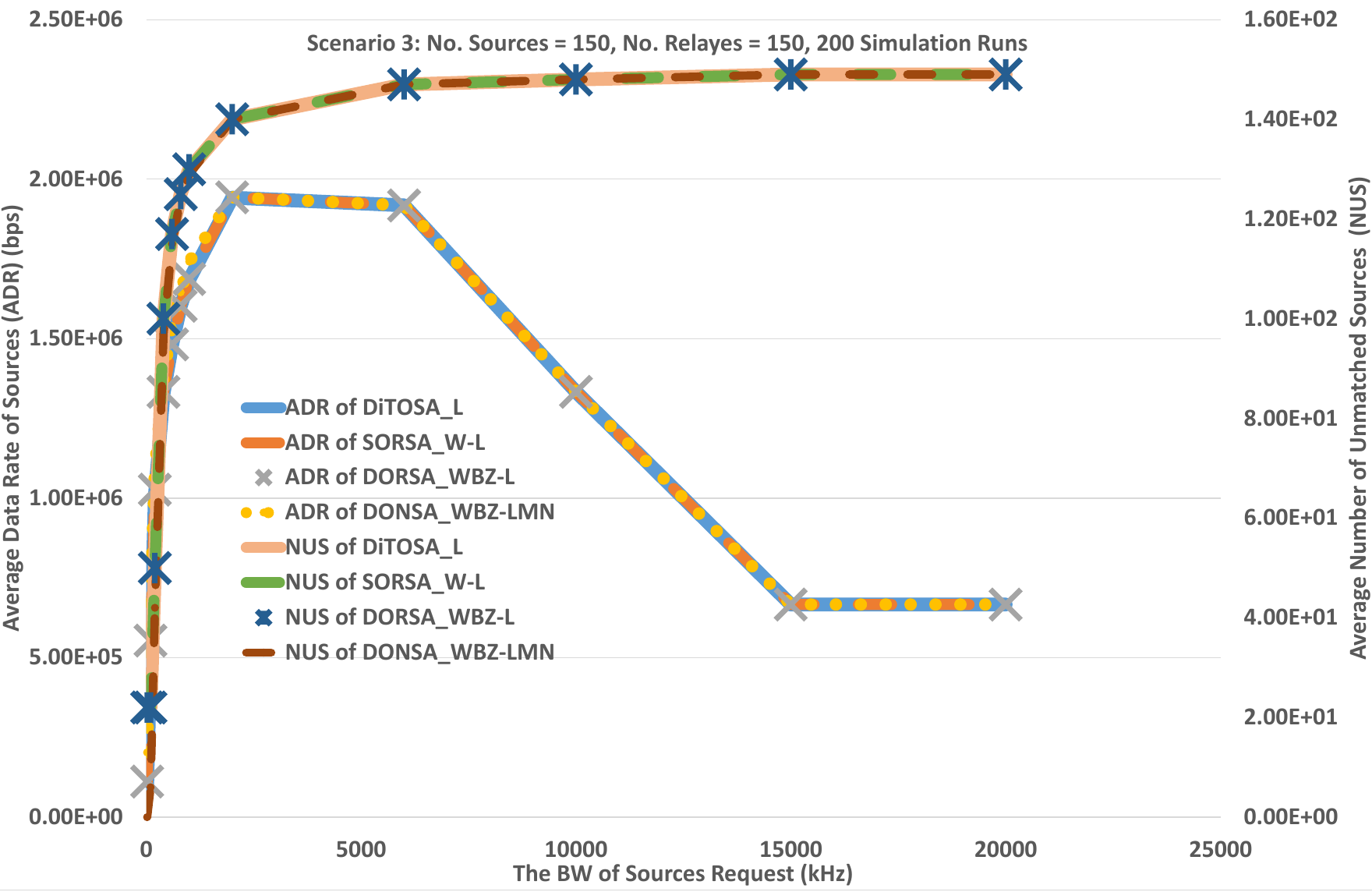}
        \label{fig:s5-3_ADR_NUS}
    }    
    \caption{ADR and NUS vs. No. sources, cell radius, and the requested BW of sources in different Scenarios with 200 simulation runs.}
    \label{fig:s1s4s5}
\end{figure}

In all cases of Fig.~\ref{fig:s1s4s5} is observed that the simultaneous use of the BW capacity of all M2M and M2B RF interfaces in DONSA\_WBZ-LMN improves the ADR and NUS compared to other algorithms. As can be seen in Fig.~\ref{fig:s1_ADR_NUS} related to Scenario 1, As the No. sources increases and the No. relays decreases, the available bandwidth is divided among a number of them. For this reason, with the increase in the No. sources and more requests than the BS capacity, BW has not been allocated to some of them, and as a result, ADR charts have a downward trend and NUS charts have an upward trend after reaching the maximum BW capacity in each algorithm. However, it is observed that DONSA has improved ADR (by an average of 13\%) and NUS (by 1.9\%) by increasing the BS connection capacity. In Scenario 2 (Fig.~\ref{fig:s4-2_ADR_NUS}), increasing the cell radius reduces the No. RFs with the required coverage range for long-distance machines. Therefore, it is only possible to use fewer RFs (such as LTE-M and NB-IoT) with lower supported data rates, and this has led to a downward trend in the ADR. In addition, since the sources requests are the same (= 200 (kHz)) and is supported by all RFs, algorithms that use only one M2M RF (LTE) will not be able to serve at least 50 sources. In contrast, with the simultaneous use of all RFs and the increase of BS connection capacity, ADR (by an average of 117\%) and NUS (by 5.3\%) have improved.  
Fig.~\ref{fig:s5-3_ADR_NUS}, related to scenario 3, shows that As long as all RFs are usable, the ADR is incremental, but from the point on requested BW = 300 MHz onwards, the No. RFs that can support requested BW decreases, and the chart trend decreases. This process continues until the end of the chart, where, according to the requested BW, a maximum of 1 source is connected to the BS. Therefore, where the requested BW varies from $20(kHz)$ to $20(MHz)$, machines are supported by 3, 2, or 1 M2B interfaces, NUS in DONSA improves by a maximum of about 14.7\% and an average of 2.9\%. Finally, the ADR obtained in DONSA improves by an average of 11.1\%. The 95\% confidence interval of ADR for DONSA graphs in three scenarios are CI[$(1.28 \times 10^6) \pm 0.09\%$], CI[$(5.24 \times 10^5) \pm 0.4\%$], and CI[$(1.23 \times 10^6) \pm 0.4\%$], respectively. The trendline of AET in DONSA for all scenarios is less than calculated TC($O(n^3)$). Maximum AET in scenarios (1, 2, and 3), are equal to 431(ms), 1.1(s), and 1.8(s), respectively.

\section{Conclusion and Future Works}
A generalized framework for dynamic optimal M2M and M2B RF interface setting and next-hop selection for sources with the similar requested BW was proposed in an IoT network. Using this framework, machines RF equipment may be used more efficiently. The simulations show that using this method improved the data rate of network sources up to 117\% in different scenarios.
The result can be used to evaluate subsequent algorithms in this field as an upper bound. In addition, in the future, our proposed algorithm can be compared with other solvers to find the optimal solution. Also, practical decentralized algorithms can be designed for dynamic RF interface setting in next-hop selection with lower time complexity in real networks. Furthermore, algorithms can be provided in situations where sources have a variety of requested BW. 

\end{document}